%% file: urumqir.tex
\input wsharvmac

 \noblackbox
\input epsf
\epsfverbosetrue
\def\epsfsize#1#2{\hsize}
\parindent =6pt

\def\B{{\cal B}}
\def\K{{\cal K}}


%
%
\def\RF#1#2{\if*#1\ref#1{#2.}\else#1\fi}
\def\NRF#1#2{\if*#1\nref#1{#2.}\fi}
\def\refdef#1#2#3{\def#1{*}\def#2{#3}}
\def\rdef#1#2#3#4#5{\refdef#1#2{#3, `#4', #5}}

%
%
\def\ts{\hskip .16667em\relax}

\def\CMP{{\it Commun.\ts Math.\ts Phys.\ts}}

\def\FAP{{\it Funct.\ts Analy.\ts Appl.\ts}}
\def\IJMP{{\it Int.\ts J.\ts Mod.\ts Phys.\ts}}

\def\JP{{\it J.\ts Phys.\ts}}

\def\NP{{\it Nucl.\ts Phys.\ts}}
\def\PL{{\it Phys.\ts Lett.\ts}}

\def\PR{{\it Phys.\ts Rev.\ts}}

\def\TMP{{\it Theor.\ts Math.\ts Phys.\ts}}

\def\Zm{Zamolodchikov}
\def\AZm{A.B. \Zm}
\def\AlZm{Al.B. \Zm}
\def\dur{H.\ts W.\ts Braden, E.\ts Corrigan, P.E. Dorey \ and R.\ts Sasaki}
\def \Tahoe{Proceedings of the NATO
 Conference on Differential Geometric Methods in Theoretical
 Physics, Lake Tahoe, USA 2-8 July 1989 (Plenum 1990)}
%
%

\refdef\rAFZa\AFZa{A.\ts E.\ts Arinshtein, V.\ts A.\ts Fateev and
 \AZm, \lq Quantum S-matrix of the 1+1 dimensional Toda chain',
 \PL {\bf B87} (1979) 389-392}

\rdef\rBa\Ba{P. Bowcock}
{Classical backgrounds and scattering for affine Toda theory on a half line}
{DTP-96-37; hep-th/9609233}

\rdef\rBCDRa\BCDRa{P. Bowcock, E. Corrigan, P.E. Dorey and R. H. Rietdijk}
{Classically integrable boundary conditions for affine Toda field theories}
{\NP {\bf B445} (1995) 469}

\rdef\rBCRa\BCRa{P. Bowcock, E. Corrigan and R. H. Rietdijk}
{Background field boundary conditions for affine Toda field theories}
{\NP {\bf B465} (1996) 350}

\refdef\rBCDSb\BCDSb{\dur, `Aspects of perturbed conformal field
theory, affine Toda
field theory and exact S-matrices', \Tahoe}

\refdef\rBCDSf\BCDSf{\dur, `Affine Toda field theory: S-matrix versus
perturbation', Springer  Lecture Notes in Mathematics
{\bf 1510} (1992) 398}

\rdef\rBCDSc\BCDSc{\dur}
{Affine Toda field theory and exact S-matrices}
{\NP {\bf B338} (1990) 689}

\refdef\rBCDSe\BCDSe{\dur,
\lq Multiple poles and other features of affine Toda field theory',
\NP {\bf B356} (1991) 469-98}

\refdef\rCb\Cb{S. Coleman, {\it Aspects of Symmetry}, (Cambridge University
 Press 1985), p 185}

\refdef\rCd\Cd{S. Coleman, \PR {\bf D11} (1975) 2088}

\rdef\rCe\Ce{E. Corrigan}{
{\it Recent developments in affine Toda field theory},
lectures at the  CRM-CAP Summer School, Banff 1994}
{DTP-94/55; hep-th/9412213}

\refdef\rCHa\CHa{E. Corrigan and U.K.F. Harder, in preparation}

\refdef\rHh\Hh{U.K.F. Harder, University of Durham PhD Thesis}

\rdef\rCo\Co{I.\ts V.\ts Cherednik}
{Factorizing particles on a half line and root systems}
{\TMP {\bf 61} (1984) 977}

\refdef\rCMa\CMa{P.\ts Christe and G.\ts Mussardo, `Integrable
systems away from criticality: the Toda field theory and S-matrix of
the tri-critical Ising model', {\it Nucl. Phys}.
 {\bf B330} (1990) 465}

\refdef\rCMb\CMb{P.\ts Christe and G.\ts Mussardo,
\lq Elastic S-matrices in (1+1) dimensions and Toda field theories',
 {\it Int.~J.~Mod.~Phys.}~{\bf A5} (1990) 4581-4628}

\rdef\rCDRa\CDRa{E.\ts Corrigan, P.E. Dorey, R.H.\ts Rietdijk}
{Aspects of affine  Toda field theory on a half line}
{Suppl. Prog. Theor. Phys. {\bf 118} (1995) 143}

\rdef\rCDRSa\CDRSa{E.\ts Corrigan, P.E. Dorey, R.H.\ts Rietdijk and R.\ts
Sasaki}
{Affine Toda field theory on a half line}
{\PL {\bf B333} (1994) 83}

\refdef\rCDSa\CDSa{E. Corrigan, P.E. Dorey and R. Sasaki,
 \lq On a generalised bootstrap principle',
\NP {\bf B408} (1993) 579-99}

\refdef\rDGZa\DGZa{G.W. Delius, M.T. Grisaru and D. Zanon,
\lq Exact S-matrices for non simply-laced affine Toda theories',
\NP {\bf B382}  (1992) 365-408}

\refdef\rDc\Dc{P.\ts E.\ts Dorey,
\lq Root systems and purely elastic S-matrices, I',
\NP {\bf B358} (1991) 654-676}

\refdef\rDd\Dd{P.\ts E.\ts Dorey, \lq Root systems and purely elastic
S-matrices II', \NP {\bf B374} (1992) 741}

\refdef\rFb\Fb{M.\ts D.\ts Freeman,
\lq On the mass spectrum of affine Toda field theory',
\PL \hfill\break {\bf B261} (1991) 57-61}

\refdef\rFc\Fc{M.\ts D.\ts Freeman,,
\lq Conserved charges and soliton solutions in affine Toda theory',
\NP {\bf B433} (1995) 657}

\refdef\rFLOa\FLOa{A. Fring, H.C. Liao and D.I. Olive,
\lq The mass spectrum and couplings in affine Toda field theories',
\PL {\bf 266B} (1991) 82-86}

\rdef\rFKc\FKc{A.\ts Fring and R.\ts K\"oberle}
{Factorized scattering in the presence of reflecting boundaries}
{\NP {\bf B421} (1994) 159}

\rdef\rFKd\FKd{A.\ts Fring and R.\ts K\"oberle}
{Affine Toda field theory in the presence of reflecting boundaries}
{\NP {\bf B419} (1994) 647}

\rdef\rFKe\FKe{A.\ts Fring and R.\ts K\"oberle}
{Boundary bound states in affine Toda field theory}
{\IJMP {\bf A10} (1995) 739}

\refdef\rFOa\FOa{A. Fring and D.I. Olive, \lq The fusing rule and
scattering matrix of affine Toda theory',
\NP {\bf B379} (1992) 429-47}

\rdef\rFSa\FSa{A. Fujii and R. Sasaki}
{Boundary effects in integrable field theory on a half line}
{Prog. Theor. Phys. {\bf 93} (1995) 1123}

\refdef\rGc\Gc{G.M. Gandenberger, \lq Exact S-matrices for bound states of
$a_2^{(1)}$ affine Toda solitons', \NP {\bf B449} (1995) 375-405}

\refdef\rGMa\GMa{G.M. Gandenberger and N.J. MacKay, `Exact S matrices
for $d_{(n+1)}^{(2)}$ affine Toda solitons and their bound states',
\NP {\bf B457} (1995) 240}

\refdef\rGMWa\GMWa{G.M. Gandenberger, N.J. Mackay and G.M.T.
Watts, \lq Twisted algebra R-matrices and S-matrices for
$b_n^{(1)}$ affine Toda solitons and their bound states',
\NP {\bf B465} (1996) 329}

\refdef\rGd\Gd{S.\ts Ghoshal,
`Boundary state boundary S-matrix of the sine-Gordon model',
{\it Int. J. Mod. Phys.} {\bf A9} (1994) 4801}

\rdef\rHa\Ha{T.J. Hollowood}
{Solitons in affine Toda field theories}
{\NP {\bf 384} (1992) 523}

\refdef\rHf\Hf{T.J. Hollowood,
\lq Quantum soliton mass corrections in $SL(N)$ affine
Toda field theory',
\PL {\bf B300} (1993) 73-83}

\refdef\rHb\Hb{T.J. Hollowood,
`Quantizing SL(N) solitons and the
Hecke algebra', {\it Int. J. Mod. Phys.} {\bf A8} (1993) 947}

\refdef\rHg\Hg{T.J. Hollowood, `The Analytic structure of
trigonometric S matrices',  Nucl. Phys. {\bf B414} (1994) 379}

\refdef\rCKa\CKa{H.S. Cho and J.D. Kim, \lq Boundary reflection matrix for
$d_4^{(1)}$ affine Toda field theory', DTP-95-23; hep-th/9505138}

\rdef\rFSWa\FSWa{P. Fendley, H. Saleur and N.P. Warner}
{Exact solutions of a massless scalar field with a relevant boundary
interaction}
{\NP {\bf B430} (1994) 577}

\refdef\rKc\Kc{V.\ts Ka\v c, {\it Infinite Dimensional Lie Algebras}
(Birkhauser 1983)}

\rdef\rKe\Ke{J.D. Kim}
{Boundary reflection matrix in perturbative quantum field theory}
{\PL {\bf B353} (1995) 213}

\rdef\rKf\Kf{J.D. Kim}
{Boundary reflection matrix for A-D-E affine Toda field theory}
{DTP-95-31; hep-th/9506031}

\rdef\rKKa\KKa{J.D. Kim and I.G. Koh}
{Square root singularity in boundary reflection matrix}
{KAIST-THP-96-702; hep-th/9605061}

\rdef\rLa\La{G.L. Lamb, Jr}
{Elements of Soliton Theory}
{John Wiley and Sons Inc. 1980}

\rdef\rMv\Mv{A. MacIntyre}
{Integrable boundary conditions for classical sine-Gordon theory}
{{\it  J. Phys.} {\bf A28} (1995) 1089}

\rdef\rGZa\GZa{S.\ts Ghoshal and \AZm}
{Boundary $S$ matrix and boundary state in two-dimensional
integrable quantum
field theory}
{{\it Int. J. Mod. Phys.} {\bf A9} (1994) 3841}

\refdef\rJa\Ja{M.\ts Jimbo, {\it Yang-Baxter equations in integrable systems},
 (World Scientific, 1990)}

\rdef\rMw\Mw{S. Mandelstam}
 {Soliton operators for the quantized sine-Gordon equation}
{\PR {\bf D11} (1975) 3026}

\refdef\rMl\Ml{A.V. Mikhailov, `Integrability of a two-dimensional
generalisation of the Toda chain', {\it JETP Letts.} {\bf 30} (1979) 414}

\rdef\rMOPa\MOPa{A. V. Mikhailov, M. A. Olshanetsky and A. M. Perelomov}
{Two-dimensional generalised Toda lattice}
{\CMP {\bf 79} (1981) 473}

\rdef\rOTa\OTa{D. I. Olive and N. Turok}
{The symmetries of Dynkin diagrams and the reduction of Toda field equations}
 {{\it Nucl. Phys.} {\bf B215} (1983) 470}

\refdef\rOTb\OTb{D. I. Olive and N. Turok,
\lq The Toda lattice field theory hierarchies and zero
curvature conditions in Ka\v c-Moody algebras',
{\it Nucl. Phys.} {\bf B265} (1986) 469}

\refdef\rOTd\OTd{D. Olive and N. Turok,
\lq Local conserved densities and zero curvature conditions for
Toda lattice field theories',
 \NP {\bf B257} (1985) 277-301}

\refdef\rOTUb\OTUb{D.I. Olive, N. Turok and J.W.R. Underwood,
\lq Affine Toda solitons and vertex operators',
\NP {\bf B409} (1993) 509-546}

\refdef\rOTUa\OTUa{D.I. Olive, N. Turok and J.W.R. Underwood,
 \lq Solitons and the energy-momentum tensor for affine
Toda theory',  \NP {\bf B401} (1993) 663-97}

\refdef\rPZa\PZa{S. Penati and D. Zanon,
\lq Quantum integrability in two-dimensional sytems with boundary',
\PL {\bf B358} 63}

\refdef\rPRZa\PRZa{S. Penati, A. Refolli and D. Zanon,
\lq Quantum boundary currents for non simply-laced Toda theories',
IFUM-518-FT; hep-th/9510084}

\refdef\rPRZb\PRZb{S. Penati, A. Refolli and D. Zanon,
\lq Classical versus quantum symmetries for Toda theories with a non-trivial
boundary perturbation',
IFUM-522-FT; hep-th/9512174}

\rdef\rSSWa\SSWa{H. Saleur, S. Skorik and N.P. Warner}
{The boundary sine-Gordon theory: classical and semi-classical analysis}
{\NP {\bf B441} (1995) 421}

\rdef\rSSa\SSa{S. Skorik and H. Saleur}
{Boundary bound states and boundary bootstrap in the sine-Gordon model
with Dirichlet boundary conditions}
{USC-95-01; hep-th/9502011}

\rdef\rSk\Sk{R.\ts Sasaki}
{Reflection bootstrap equations for Toda field theory}
{in {\it Interface between Physics and Mathematics}, eds W. Nahm and J-M Shen,
(World Scientific 1994) 201}

\rdef\rSl\Sl{E.\ts K.\ts Sklyanin}
{Boundary conditions for integrable equations}
{\FAP {\bf 21} (1987) 164}

\rdef\rSm\Sm{E.\ts K.\ts Sklyanin}
{Boundary conditions for integrable quantum systems}
{\JP {\bf A21} (1988) 2375}

\rdef\rWa\Wa{G. Wilson}
{The modified Lax and and two-dimensional Toda lattice equations
associated with simple Lie algebras}
{{\it Ergod. Th. and Dynam. Sys.} {\bf 1} (1981) 361}

\rdef\rYa\Ya{A. Yegulalp}
{New boundary conformal field theories indexed by the simply-laced
Lie algebras}
{\NP {\bf B450} (1995) 641}

\rdef\rZZa\ZZa{\AZm\  and \AlZm}
{Factorized S-matrices in
two-dimensions as the exact solutions of certain relativistic
quantum field models}
{{\it Ann. Phys.} {\bf 120} (1979) 253}

\rightline{DTP-96/49}
\rightline{hep-th/9612138}
\bigskip
{\obeylines \centerline{\bf INTEGRABLE FIELD THEORY WITH}
\centerline{\bf BOUNDARY CONDITIONS}}
\vskip 3pc
\centerline{E. CORRIGAN}
\bigskip
\centerline{\it Department of Mathematical Sciences}
\centerline{\it University of Durham}
\centerline{\it Durham DH1 3LE, England}
\medskip
\vskip 1pc
\parindent=40pt
{\narrower\smallskip\noindent Talks given at the workshop {\it Frontiers in
Quantum Field
Theory}, held in Urumqi, Xinjiang Uygur Autonomous Region, People's
Republic of China, 11-19 August 1996.\smallskip}
\parindent=10pt

\newsec{Introduction}

The purpose of these talks is to  review some of the ideas
surrounding the topic of
integrability, either classical or quantum, when two-dimensional
field theory (i.e. one space---one time)
is restricted to a half-line, or to a segment of the
line.

There are some simple questions which may be asked but they do
not turn out to have simple answers. Moreover, there are some surprises
and unexpected constraints once particular models are examined.
The affine Toda field theories, or $\sigma$-models provide excellent
illustrations.

For example, suppose an integrable field theory describes a
collection of
distinguishable particles (real affine Toda field theory has this
property---the scalar particles are distinguished by conserved charges
of non-zero spin), what is the spectrum of particle energies if
the system is enclosed on the interval $[-L,L]$?

Alternatively, if the theory is confined to the space region $x<0$,
and a particle approaches $x=0$ from the left one might expect the
minimal effect of the boundary to be a reversal of all the
momentum-like conserved quantities and the preservation of all energy-like
conserved quantities. If that is the case, the particle preserves its
identity
but reverses its direction. It might be expected that the \lq out' state
consisting of a single particle is proportional to the \lq in' state
with the momentum reversed. In other words:
\eqn\Kfactor{|k>_{\rm out}=K_a(k)|-k>_{\rm in}}
where $a$ is a particle label. It might happen (for example in the
$\sigma$-models) that the particles belong to multiplets whose members are
distinguishable only via spin-zero charges. In those cases, the reflection
factor
$K_a$ appearing in Eq.\Kfactor\ will be a matrix to allow for
a mixing of states at the boundary. Labels corresponding to multiplicity
will generally  be suppressed
unless there is a specific reason to display them.

If the particles of an integrable theory are distinguishable scalars
then the possible momenta for a particle of type $a$ confined to the interval
$[-L,L]$ might be expected to be given by the solutions to
\eqn\momenta{e^{4ikL}K_a^{(L)}(k)K_a^{(-L)}(-k)=1}
where the superscripts on the reflection factors take into account
possibly different boundary conditions at the two ends of the interval.
Eq.\momenta\ may be a false assumption but it is based on the idea
that in an integrable theory factorization is a fundamental property,
and the effects of the two boundaries ought to be independent of
each other. It is clearly true for a particle described by a free field
with linear boundary conditions
\eqn\linear{\partial_x\phi|_{x=\pm L}=a_{\pm}\phi|_{\pm L}.}
But otherwise, of course, such a claim requires proof.
If it is correct
then Eq.\momenta\ provides a motivation for the study of the reflection
factors themselves, even in the simplest of models.

Much is now known about classical and quantum integrability
for field theories defined over the whole line in a two-dimensional
space-time. Several classes of models (eg affine Toda theory,
or $\sigma$-models) have been particularly well studied, as has the
sine-Gordon model. The latter is essentially the $a_1^{(1)}$ affine
Toda theory with a purely imaginary field.
On the other hand, the situation in which the
line is truncated to a half-line, or to an interval, is hardly explored
except for the choice of Neumann, or periodic boundary conditions.
In particular, until recently, it was not known what boundary conditions
were compatible with integrability, even classically.

There has been renewed interest in this topic following
investigations in condensed matter physics in which boundaries
play a significant r\^ole. In particular, the sine-Gordon model
has been studied afresh, with a number of new results obtained,
notably by Ghoshal and
Zamolodchikov,\NRF\rGZa\GZa\NRF\rGd\Gd\refs{\rGZa ,\rGd} and others
\NRF\rSSWa{\SSWa\semi\SSa}\refs{\rSSWa}
(see Fendley, Saleur and Warner
\NRF\rFSWa\FSWa\refs{\rFSWa}). The sine-Gordon
model is the simplest of the affine Toda field theories, based
on the root data of the Lie algebra $a_1$, and it is therefore natural
to explore the question of boundary conditions within the context of
other  models in the
same class.\NRF\rFKc{\FKc\semi\FKd}\NRF\rSk\Sk\refs{\rFKc , \rSk}

However, before doing so, it is worthwhile to review and explore some
of the properties which are expected to hold for the reflection factors in
quantum field theory and which were conjectured
some time ago by Cherednik.\NRF\rCo\Co\refs{\rCo}

\newsec{Quantum integrability: conjectures}

The question of quantum integrability in the presence of a boundary is
difficult to tackle. It appears the best one can hope to do at present
is to set out
a set of hypotheses which at least allow consistent conjectures to be made
which may themselves be checked subsequently in various ways. For example,
in certain circumstances perturbation theory might be relevant.
The principal ideas were set
out by Cherednik
\NRF\rCo\Co\refs{\rCo}
many years ago and have been supplemented recently in work of Ghoshal and
Zamolodchikov
\refs{\rGZa} concerning the sine-Gordon theory, and by
Fring and K\"oberle,\NRF\rFKc{\FKc\semi\FKd}\refs{\rFKc} and by Sasaki
\NRF\rSk\Sk\refs{\rSk} concerning the affine Toda theories. Consider
first the situation with a single boundary.

The principal idea is that particle states in the presence of the boundary
(taken to be at
$x=0$)
continue to be eigenstates of energy and other energy-like
conserved charges. However, an
initial state containing a single particle moving towards the boundary will
evolve into
a final state with a single particle moving away from the boundary. Thus
\eqn\quantumK{|a, v>_{\rm out}=K_{ab}(\theta )|b, -v>_{\rm in},}
where the states $a,b$ correspond to multiplets of particles distinguishable
merely via spin-zero charges, and $K_{ab}$ is a matrix which may mix the
particles
as a result of the reflection from the boundary.
For real affine Toda theory
 the particles are all distinguishable and therefore there should be a set
of reflection factors, one for each particle, corresponding to each integrable
boundary
condition. The velocity of  a particle is reversed on reflection, and this is
equivalent to reversing its rapidity (defined by $v=\tanh\theta$). The first
major
task is to determine a set of $K$ factors for a specified boundary
condition.

This situation is represented pictorially by:
\vskip -10 cm
\epsffile{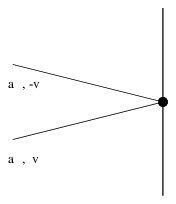}
\noindent where the vertical line represents the world-line of the boundary at
$x=0$.

It is also supposed that particles scatter factorizably, but
independently of the
boundary, when they are far from it. As mentioned before, such
an assumption requires proof within any particular model. It is a plausible
assumption since the boundary would not be expected to to affect particles
moving towards it until they were `close' by. However, if
it is true then the usual arguments
\NRF\rZZa\ZZa\refs{\rZZa} require the S-matrix
describing the scattering of two-particle states:
\eqn\Smatrix{|\theta_a,\theta_b>_{\rm
out}=S_{ab}(\theta_a-\theta_b)|\theta_a,\theta_b>_{\rm in}}
to satisfy a Yang-Baxter equation
\eqn\YB{S_{ab}(\theta_a -\theta_b )S_{ac}(\theta_a -\theta_c )S_{bc}
(\theta_b -\theta_c )=S_{bc}(\theta_b -\theta_c )S_{ac}
(\theta_a -\theta_c )S_{ab}(\theta_a -\theta_b ).}
As in Eq.\quantumK , the subscripts denote particle types, the matrix labels
have been omitted deliberately.

Eq.\YB\  may be represented pictorially by:
\vfill\eject

\epsffile{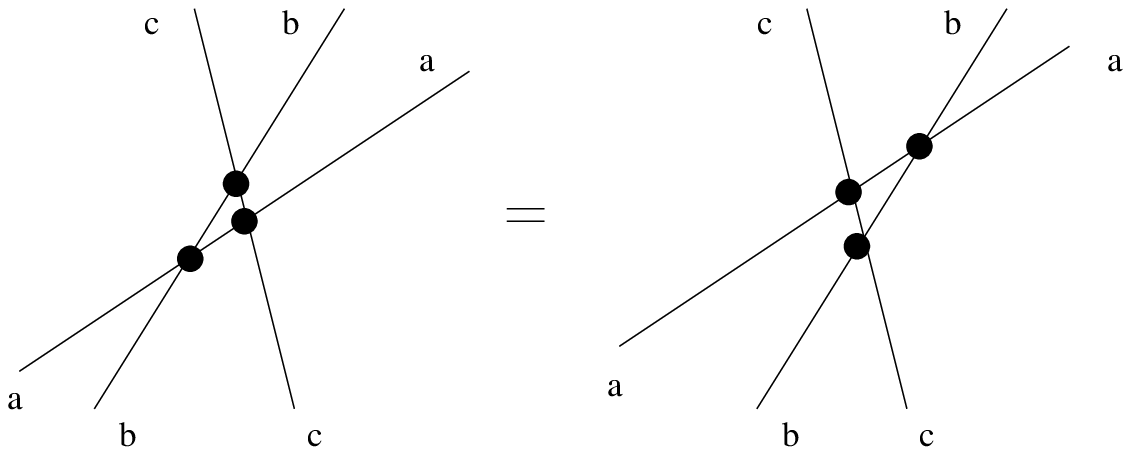}

This is something of a constraint unless the S-matrix factors are merely
sets of phases (as they are in a situation where the particles are
distinguishable), but it fails to determine the S-matrix completely. If
the particles are distinguishable, Eq.\YB\ is an identity. The
general solution
to the Yang-Baxter equation is not known although
general classes of
solution have been found, including those related to the theory of quantum
groups (see for instance the reprint
volume by Jimbo
\NRF\rJa\Ja\refs{\rJa}).

On the other hand, a two-particle state consisting of two incoming particles
will eventually evolve into a state containing the two outgoing particles.
However, each of the particles will scatter from the boundary and,
inevitably from each other. But, the order of the individual
scatterings and reflections should not matter because they
depend (supposedly inessentially) on the initial condition
setting up the two-particle state. If it is supposed that these events
also take place factorisably then one obtains the
reflection Yang-Baxter equation.
 Algebraically, the relationship is
\eqn\boundaryYB{K_a(\theta_a )S_{ab}(\theta_b+\theta_a )K_b(\theta_b
)S_{ab}(\theta_b-\theta_a)=
S_{ab}(\theta_b-\theta_a )K_b(\theta_b )S_{ab}(\theta_b+\theta_a)
K_a(\theta_a
).}
For affine Toda field theory where the particles are distinguishable,
$K$ and $S$ are diagonal, and the boundary Yang-Baxter equation is
satisfied identically.

Pictorially, Eq.\boundaryYB\  would be represented by:

\epsfxsize=4cm
\epsfysize=5.5cm
\epsffile{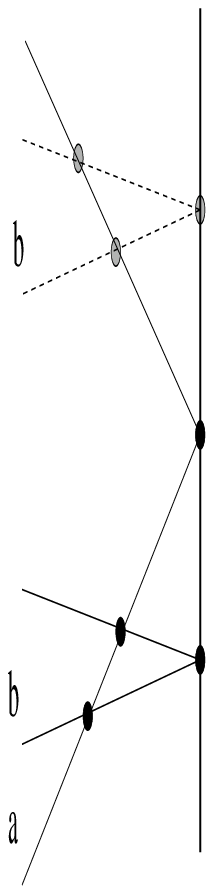}

For the whole line theory there is a consistent bootstrap principle,
in the
sense that
there is a consistent set of couplings between the particles,
signalled by the
presence
of poles in the S-matrix at certain (imaginary)
relative rapidities, and these
may be
used to relate the S-matrix elements to each other.
The set of bootstrap relations take the form
\eqn\bootstrap{ S_{dc}(\theta_d -\theta_c)
=S_{da}(\theta_d-\theta_a)S_{db}(\theta_d-\theta_b)}
where
\eqn\thetaboot{\theta_a=\theta_c-i\bar\theta_{ac}^b\qquad
\theta_b=\theta_c+i\bar\theta_{bc}^a,}
$\bar\theta =\pi -\theta$, and the coupling angles are the angles of the
triangle
with side-lengths equal to the masses of particles $a,b$ and $c$ participating
in the coupling $ab\rightarrow \bar c$.
Not every pole
in the S-matrix factors signal the existence of a bound state, however. In
the real affine Toda theories corresponding to data for the $a^{(1)},d^{(1)}$
or $e^{(1)}$ series of affine root systems, the relevant poles are those of odd
order with a coefficient of the correct sign
\NRF\rBCDSc{\BCDSc\semi\BCDSe}\refs{\rBCDSc}. Moreover, the couplings
themselves follow Dorey's rule
\NRF\rDc{\Dc\semi\Dd\semi\Fb\semi\FLOa}\refs{\rDc}, which is intimately related
to the properties of root systems and the Coxeter element of the Weyl group.
For the other theories, based
on affine root data corresponding to the non simply-laced diagrams, the
situation is rather different. These cases, demonstrate some remarkable
phenomena in the sense that the quantum bound states do not appear
to have mass ratios equal to the classical values (in the $ade$ cases they do).
Rather, the masses \lq float' between the classical values
corresponding to dual pairs (see later in section(3)). The details may be
found in Delius et al. and elsewhere
\NRF\rDGZa{\DGZa\semi\CDSa}\refs{\rDGZa}. For these,
the geometrical setting for the couplings (that is,
a suitable generalization of Dorey's rule)
is not known.

\epsffile{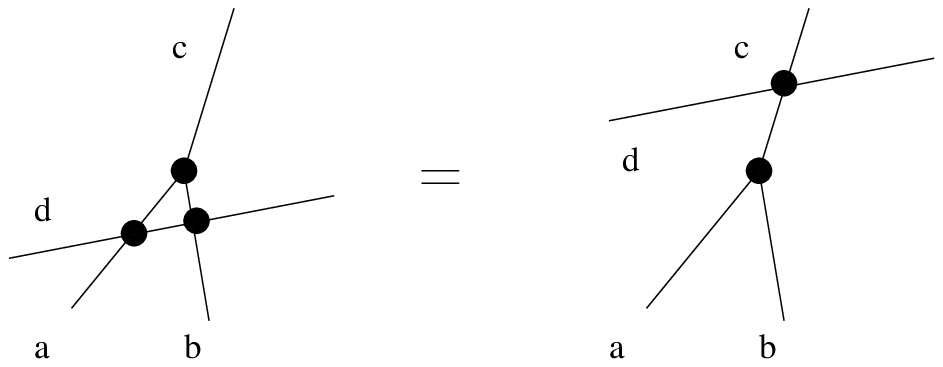}

Assuming  the family of whole line
couplings
remains relevant in the presence of a boundary, the bootstrap implies relations
between the various reflection factors. At the special imaginary relative
rapidities corresponding to a bound state, the two particle $a,b$ states have
quantum numbers identically equal to the quantum numbers of particle $c$.
One might think picturesquely of either $a,b$ separately reflecting from the
boundary in advance (or after) the bound state forms, or the particle $c$
reflects
from the boundary.
Algebraically, the reflection bootstrap
equation is:
\eqn\Rbootstrap{K_c(\theta_c)=K_a(\theta_a)S_{ab}(\theta_b+
\theta_a)K_b(\theta_b),}
where $\theta_a,\theta_b$ are given in Eq.\thetaboot .
\vfill\eject
This too has a pictorial representation:

\vskip -2cm
\epsffile{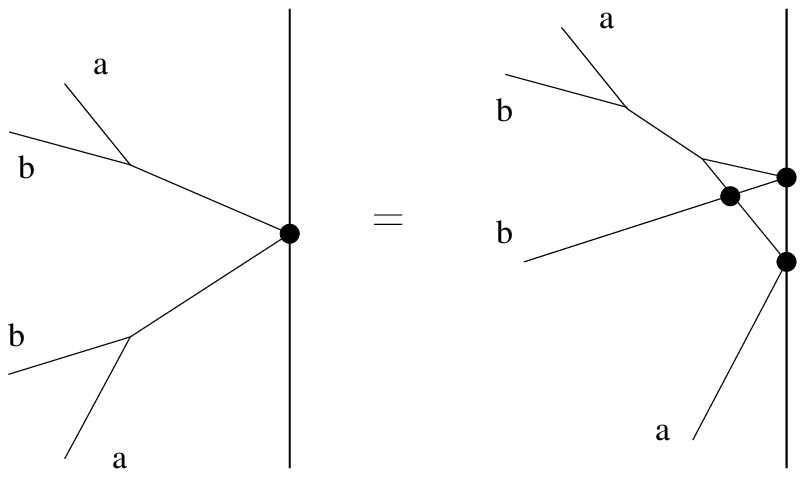}

There is also the possibility of  bound states involving a particle and the
boundary---with their own coupling angles and bootstrap property
(see\NRF\rCDRSa\CDRSa
\NRF\rFKe\FKe\refs{\rGZa ,\rCDRSa , \rFKe}). The idea is that the
boundary can be excited, and that this possibility should be indicated by the
existence of
poles in $K_a(\theta_a )$. For example, particle $a$ and boundary-type
$\alpha$ might have a reflection factor $K_a^\alpha$ with a suitable
pole at
$\theta_a =i\psi_{a\alpha}^\beta$, which is to be interpreted as a signal
for a boundary bound state labelled $\beta$, say. One would expect
the \lq physical' strip for $\psi_{a\alpha}^\beta$ to lie in the
range $[0,\pi /2]$. A two-particle state consisting of $a$ and another
particle $b$ would also encode the pole in $\theta_a$, in which case the
the particle $b$ should be regarded as either reflecting from the boundary
state $\beta$
directly, or, alternatively, reflecting from the boundary state
$\alpha$. In the latter case it scatters twice with particle $a$,
once before,
and once after its reflection from the boundary state $\alpha$. Of
course, such an interpretation also relies heavily on the
factorisation assumption. Algebraically, one would write,
\eqn\bbound{K_b^\beta (\theta_b )=S_{ab}(\theta_b+i\psi_{a\alpha}^\beta )
K_b^\alpha (\theta_b)S_{ab}(\theta_b-i\psi_{a\alpha}^\beta ).}
Eq.\bbound\ gives a set of consistency conditions which would need to
be satisfied by the boundary bound states and their associated
reflection factors. Almost no work has been done to determine the
rules corresponding to the boundary bound state poles. Nor has there been
a systematic determination of the full set of reflection factors for
any specific model. It might be imagined that one could begin
with a full set of ground state factors and simply calculate the rest
self-consistently. However, that turns out to be easier said than done
because it has simply proved impossible, so far, properly  to analyse the
pole structure of the reflection factors. It has even been suggested that
the reflection factors might have singularities of square-root type
\NRF\rKKa\KKa\refs{\rKKa}. If that is genuinely so, then the whole question
of the analytic structure needs to be re-examined. Even a free field with a
linear boundary condition of the type given in Eq.\linear , has a reflection
factor of the form
\eqn\linearK{K(k)={ik-a\over ik+a}\ ,}
containing  a pole at $k=ia$. This pole corresponds to an exponentially
decreasing wave function in the region $x<0$, provided $a$ is positive,
and therefore a normalizable solution to the field equation representing
a bound state. Some ideas concerning boundary bound states
within the sine-Gordon model
are to be found in
Ghoshal and Zamolodchikov's paper
\NRF\rGZa\GZa\refs{\rGZa}, while some conjectures concerning the affine Toda
theories are to be found elsewhere
\NRF\rCDRSa\CDRSa\NRF\rFKe\FKe\refs{\rCDRSa ,\rFKe}.

Pictorially, the boundary bound state bootstrap would be represented by
something like:
\vskip -2cm
{

\epsfbox{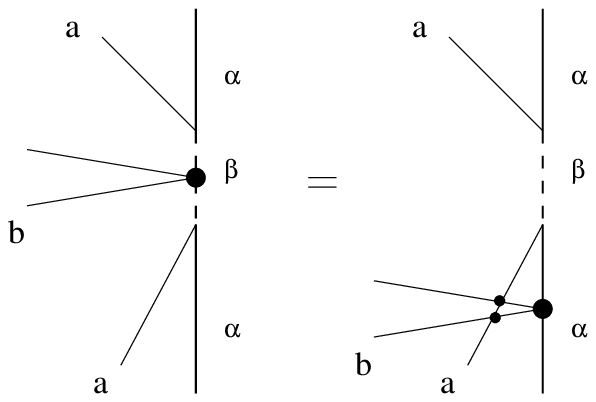}}

Finally, there are the Crossing relations
\eqn\Scrossing{S_{ab}(i\pi -\Theta )=S_{a\bar b}(\Theta )=
S_{\bar a b}(\Theta )}
where $\Theta =\theta_a-\theta_b$,
\eqn\Kcrossing{K_a(\theta-i\pi /2)K_{\bar a}(\theta+i\pi /2)
S_{\bar a a}(2\theta )=1;}
and the Unitarity relations
\eqn\SKunitarity{S_{ab}(\Theta )=S^{-1}_{ab}(-\Theta )\qquad K_a(\theta
)=K_a^{-1}(-\theta
).}
These are written in a manner appropriate to a model in which all particles
are distinguishable (as were the bootstrap relations). If the particles
belong to multiplets these relations need to be suitably modified; in the
cases of Eqs.\Kcrossing , \Scrossing\ a crossing matrix is needed.

Even for a collection of relatively simple models such as affine Toda theory,
there are many known solutions to the reflection bootstrap equations
\refs{\rSk}. However, it
is not clear how to relate them to the possible choices of
boundary condition.
Presumably, it will be necessary to apply a semi-classical approximation,
or to use perturbation theory, although the latter may be difficult in
situations
where there are no small parameters associated with the boundary potential.
There has been some work in this direction by Kim
\NRF\rKe{\Ke\semi\CKa\semi\Kf}\refs{\rKe}, but only concerning the Neumann
boundary condition.
Quantum versions of the conserved quantities have also been investigated
recently by Penati and Zanon
\NRF\rPZa\PZa
\NRF\rPRZa{\PRZa}\refs{\rPZa ,\rPRZa} whose calculations suggest
that the boundary parameters will require renormalisation. It is
notable, however, that at least in the $a_r^{(1)}$ series of cases the boundary
parameters renormalise together
\NRF\rPRZa{\PRZa}\refs{\rPZa}. However, it will be seen later on that the
permissable boundary conditions, compatible with integrability are not
usually continuously connected to the Neumann condition. Rather, the
integrable boundary conditions are \lq isolated'. For this reason,
perturbation
using a boundary parameter ought probably to be treated
with great caution.
\medskip
\noindent{\bf A classical bootstrap}
\medskip

One interesting fact is the following
\NRF\rCDRa\CDRa\refs{\rCDRa}. If the classical limit is taken in Eq.\Rbootstrap
then it is certainly expected that the S-matrix becomes unity. However, in the
presence of
a boundary, the corresponding limit of the reflection factors need
not necessarily be
unity; remember, even a free particle must rebound from the boundary,
and a linear boundary condition of the type given in Eq.\linear\
leads to a reflection factor of the form given in Eq.\linearK .
Rather, the classical limit $K_0$ might be expected to satisfy the classical
limit
of the reflection bootstrap equation \Rbootstrap . Moreover, these classical
limits are
themselves computable via a standard linearised scattering problem involving
an \lq effective potential' determined by the (presumably static)
lowest energy solution to the classical field equations. The quantum theory
would
have to be constructed in terms of perturbations around the basic solution but
perhaps surprisingly the
classical problem already has a rich structure (including, in some cases bound
states)
\refs{\rCDRSa ,\rCDRa}. Some other comments
on this idea will be made later, in section(5), and further details and
developments
may be found in a recent paper by Bowcock
\NRF\rBa\Ba\refs{\rBa}.
\medskip
\noindent{\bf A strategy}
\medskip
Thus, the following strategy suggests itself. First, determine the possible
boundary conditions compatible with integrability, at least in
the sense
of maintaining as many of the desirable features usually
associated with integrability on the whole line, such as
an infinity of conserved quantities in involution. Second, determine the
classical reflection factors associated with each of the allowed
possibilities. Third, use these classical limits as a
guide to guess
suitable solutions to all the above consistency relations. Finally, check
the hypothetical reflection factors against perturbation theory.

The first part of this programme can be  carried through fully for the
affine Toda theories and that work forms the principal content of the next
sections. The second part has been carried out in many examples and it
is fascinating to see the classical limit of the bootstrap relation
actually working in practice for the \lq classical' reflection factors.
Often these rely on wonderful identities. Ultimately, though, the
beautiful structure can be traced to the properties of what might be
called \lq solitons' in the real affine Toda field theories. These will
be discussed briefly in section (5).

On the other hand, it must be said that although this is
a clear enough strategy, there are a number of serious obstacles in the way
of carrying it out fully. In particular, the final part of the programme
needs to be tackled and the necessary perturbation theory has been
neglected up to now.
\newsec{The affine Toda theories}

The affine Toda field theories (for a recent review, see
\NRF\rCe\Ce\refs{\rCe}) are scalar theories with Lagrangian
\eqn\Ltoda{{\cal L}_{0}={1\over 2}\partial_\mu\phi\cdot\partial^\mu\phi
-{m^2\over \beta^2}\sum_{i=0}^r n_i\, e^{\beta \alpha_i\cdot\phi}.}
In this Lagrangian $m$ and $\beta$ are two constants (which may be removed
from the classical field equations by a rescaling of the fields
and the space-time coordinates); the important information
is carried by the set of vectors $\alpha_i$ and the set of
integers $n_i$.

The vectors $\alpha_i,\ i=1,2,\dots ,r$ are a set
of simple roots for a Lie algebra $g$, meaning that they
are linearly independent and any other root
may be expressed as an integer linear combination of these with either
all coefficients positive, or all coefficients negative. In particular,
the special root $\alpha_0$ is a linear combination of the
simple roots,
$$\alpha_0=-\sum_1^rn_i\alpha_i$$
where the choice of coefficients depends on $g$. For the $ade$
series of cases, $\alpha_0$ is always the \lq lowest' root. In terms of the
extended Dynkin diagrams classifying Kac-Moody algebras, $\alpha_0$
is  the Euclidean part of the extra, \lq affine' root. For the full list
of extended Dynkin diagrams, see the book by Kac
\NRF\rKc\Kc\refs{\rKc}. The diagrams full into two
classes: there are the $a,d,e$-type (including $a_{2n}^{(2)}$) which are
symmetric under the root transformation
\eqn\invert{\alpha\rightarrow 2{\alpha\over\phantom{^2} |\alpha|^2};}
and the others which come in pairs related to each other by \invert .
The first type are summarised in the first diagram below, in which the circle
representing $\alpha_0$ is
indicated by a $0$, and the simple roots are represented by circles
labelled by the integers $n_i$. The second type are listed in the
subsequent set of diagrams, in  pairs.

The classical field theories are integrable in all cases. Each of the theories
may be described by a Lax pair (the details will appear in the next section)
and, as a consequence each theory has a characteristic set of conserved
quantities: infinitely many, in fact, in involution with one another. The
classical
theory was developed over a number of years and by many authors
\NRF\rMl{\Ml\semi\MOPa\semi\Wa\semi\OTa\semi\OTb\semi\OTd}\refs{\rMl}.
More recent developments building on the older work,
including a description of the field theory
data (masses and couplings) are described in the author's Banff lecture notes
\NRF\rCe\Ce\refs{\rCe} where a fuller set of references may be found.

As far as the whole line is concerned, the quantum field theory of the
real affine Toda field theory was studied first by Arinshtein, Fateev and
Zamolodchikov
\NRF\rAFZa\AFZa\refs{\rAFZa}, for the case corresponding to the data for the
affine algebra $a_n^{(1)}$. They discovered the remarkable fact that
the quantum spectrum of states is essentially described by the classical
parameters. Their result was generalised eventually to include all the sets of
affine root data, but the generalisation turns out to be not entirely
straightforward.

The two classes, self-dual and dual pairs, behave differently after
quantisation. As far as the dual pairs are concerned the behaviour is
a generalisation of the results obtained by Arinshtein, Fateev and
Zamolodchikov, although much more intricate, and has been developed in a number
of papers
\NRF\rBCDSb{\BCDSb}
\NRF\rBCDSc{\BCDSc\semi\BCDSe\semi\BCDSf}
\NRF\rCMa{\CMa\semi\CMb}\refs{\rBCDSb ,\rBCDSc ,\rCMa}.
For these cases, there is also a beautiful formula for
the S-matrix elements, due to Dorey
\NRF\rDc{\rDc\semi\rDd}\NRF\rFOa\FOa\refs{\rDc ,\rFOa}. On the other
hand, unexpectedly, the dual pairs behave very differently
\NRF\rDGZa{\DGZa\semi\CDSa}\refs{\rDGZa}. The two sets of theories exhibit a
remarkable
type of strong-weak coupling duality. However, they achieve the duality
differently. As far as the first set is concerned,
the duality is an invariance in the sense that the set of conjectured
S-matrices has a symmetry under the mapping
\eqn\invertbeta{\beta\rightarrow {4\pi \over \beta}\, .}
Moreover, for these models the quantum spectrum of states is essentially
identical with the classical spectrum in the sense that the masses
come in the same ratio before and after quantisation, and the classical
three-point couplings and the pairwise formation of bound states
are in one to one correspondence.
However, for the second set the duality is much more subtle. For these,
for each pair it seems there is a single quantum field theory, in the
sense that the S-matrix (all these hypotheses are checked only
for S-matrices) would be the same computed from
the classical data corresponding to either of the classical lagrangians
of the pair. The mapping Eq.\invertbeta\ remains a symmetry of the S-matrix
but, for small $\beta$ perturbation theory based on one
of the classical theories corresponds to the large $\beta$ limit of the other.
The mass spectrum of these models depends on the coupling and the
$\beta$ dependent masses interpolate the classical mass parameters of
the members of the pair. The classsical couplings of the two models in
each pair are not the same (even in number!) and the rules governing the
formation of bound states are extremely subtle. A more detailed account
of how this works in the simplest of cases ($g_2^{(1)}$ - $d_4^{(3)}$)
may be found elsewhere
\NRF\rCe\Ce\refs{\rCe}.

This kind of duality is not to be confused with the well-known
particle-soliton duality which is exhibited by the sine-Gordon model,
according
to which the model may be alternatively formulated completely
differently---as the massive Thirring model
\NRF\rCb{\Cd\semi\Mw\semi\Cb}\refs{\rCb}. It is not known
if the other affine Toda theories enjoy a
similar feature. They certainly have \lq solitons' in the sense of complex
solutions with localised energy, and these solitons have interesting
properties
\NRF\rHa{\Ha\semi\Hf\semi\OTUa\semi\OTUb\semi\Fc}{\refs\rHa}.
There are conjectured S-matrices for them
\NRF\rHb{\Hb\semi\Hg}\refs{\rHb} and, in some cases,
complete solutions to the bootstrap constraints
\NRF\rGc{\Gc\semi\GMa\semi\GMWa}\refs{\rGc}, including \lq breathers'.
However, the question of the existence of an alternative formulation
in which the solitons
are fundamental remains open.

It will be seen later that the two sets of theories also behave differently
when a boundary is involved. The self-dual set (with the exception of the
sine-Gordon model and the models based on data from $a_{2n}^{(2)}$) does not
appear to allow extra parameters to be introduced
via boundary conditions. On the other hand, most of the theories
corresponding to the dual-pairs allow limited freedom at the
boundary in the sense of a small number
of free parameters. In some of the latter cases there is the possibility of a
continuous deformation away from the Neumann boundary condition. For that
reason, these may be good examples within which to develop the perturbation
theory needed for checking conjectured quantum reflection factors.

\vfill\eject
\epsffile{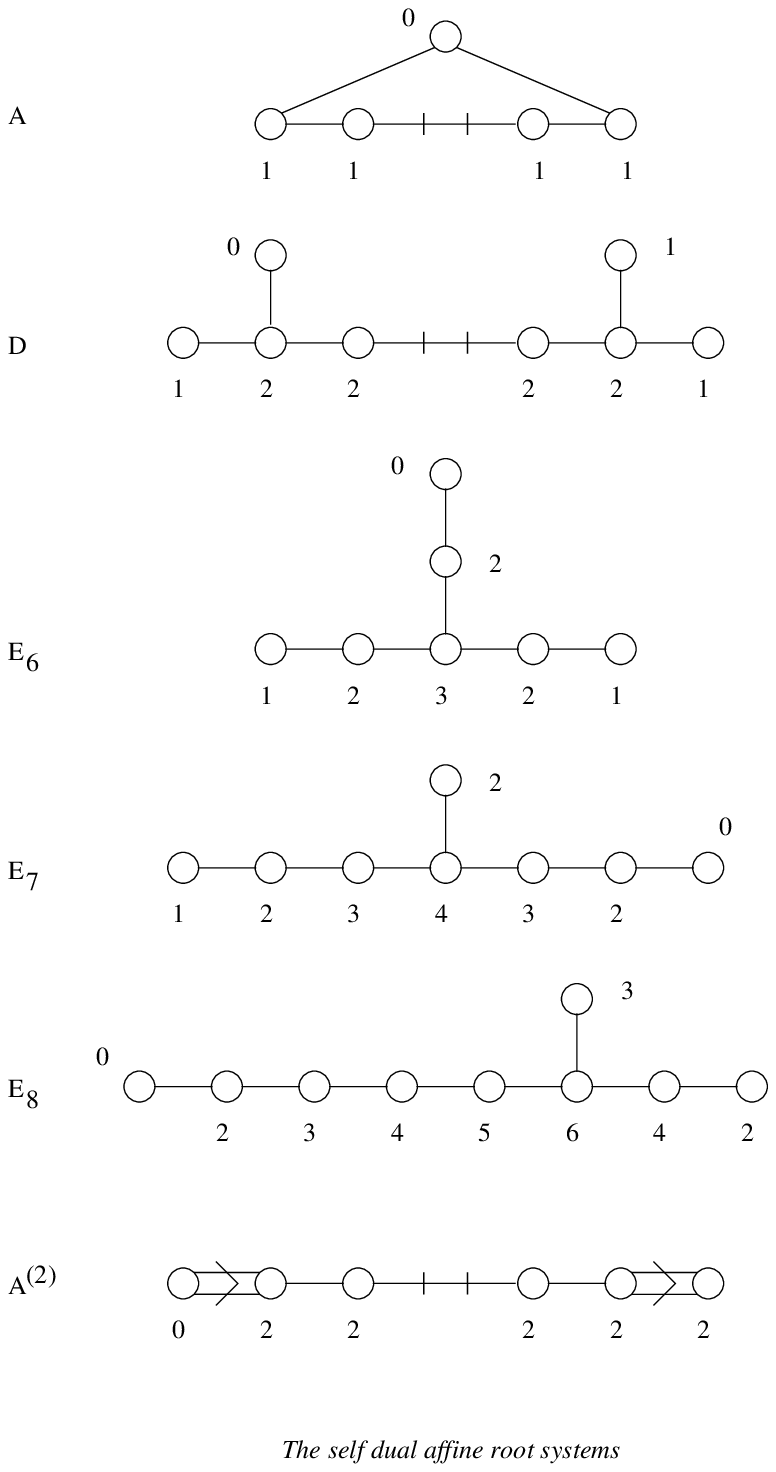}

\vfill\eject

\epsffile{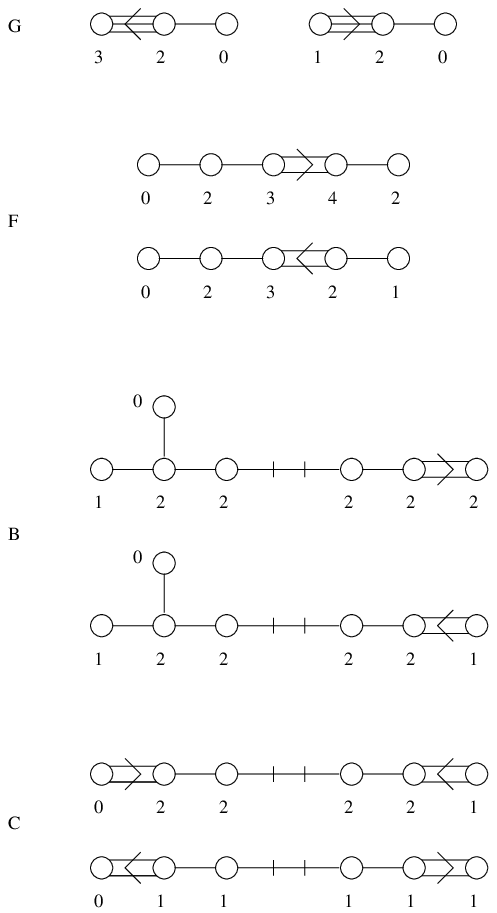}
\vskip 1 cm
\centerline{\it Dual pairs of affine root systems}

\newsec{Affine Toda theory on a half-line}

If the Toda field theory is restricted to a half-line (say,  $x\le 0$)
then there must be a boundary condition at $x=0$. In other words, the
Lagrangian
must be modified and might for example take the form:
\eqn\Lboundary{{\cal L}_{\rm B}=\theta (-x){\cal L}_0-\delta (x){\cal B}
(\phi ),}
where ${\cal L}_0$ is the Lagrangian for the whole line theory, Eq.\Ltoda ,
and it has  been assumed that the boundary term depends only on the
fields, not on their
derivatives. The latter assumption is not strictly necessary
\NRF\rBCRa\BCRa\refs{\rBCRa} but making it simplifies the discussion.

As a consequence of
\Lboundary , the field equations are restricted to the region $x\le 0$
and are supplemented by a boundary condition at $x=0$:
\eqn\equations{\eqalign{ x\le 0:\qquad \partial^2\phi_a&=-\sum_{i=0}^r
(\alpha_i)_a n_i\,
e^{\alpha_i\cdot\phi}\qquad\cr
x=0:\qquad\partial_x\phi_a&=-{\partial{\cal B}\over\partial\phi_a}\ .\cr}}
The first question to ask is the following: What choices of ${\cal B}$ are
compatible with integrability?

The question may be tackled via several routes. One way, probably the
simplest, is to consider the densities of the conserved charges which
integrate to yield the conserved charges for the full-line theory, and
discover how to modify them to preserve \lq enough' charges to maintain
integrability. Another is to develop a generalisation of the standard
Lax pair approach, including modifications arising from the boundary
condition, and to use it to investigate the charges and their
mutual Poisson brackets. The first approach is somewhat limited because
it is clearly not  feasible to study more than the first couple of
low spin conserved charges. Nevertheless, even a partial investigation
already leads to some surprising
conclusions and there is a strong suspicion that exploring the first
few conserved quantities beyond energy is probably enough for all practical
purposes. The second approach is also needed in order to be certain that
conditions found by studying low spin charges are in fact all that are
necessary. It is clearly preferable. However, a full inverse scattering
procedure in the presence of boundary conditions will be needed eventually.

The addition of  a boundary  effectively removes translational
invariance and it is no longer expected, therefore, that momentum should
be conserved. Lorentz invariance is also inevitably lost. Nevertheless, the
total energy is given by
$$\widehat E=\int_{-\infty}^0dx\ {\cal E}_0\ +\ {\cal B}(\phi_0 ),$$
where ${\cal E}_0$ is the usual energy density corresponding to ${\cal L}_0$,
and is easily seen to be conserved whatever the choice of ${\cal B}$
might be.
(The subscript on the argument of ${\cal B}$ emphasises the fact that for
this extra term the field is evaluated at $x=0$.)
These two elementary remarks already demonstrate that the best one
might hope for is
that parity even charges (like energy) might continue to be conserved
whilst parity odd charges (like momentum) cannot be.

It has already been remarked that integrable theories have
infinitely many conserved charges labelled by
their spins. For the affine Toda theories based on the root data for the
algebra $g$, the possible spins are the
exponents of the algebra modulo its Coxeter number. For example, the Lie
algebras $a_r$ has Coxeter number $h=r+1$ and the associated
affine Toda field theory has classically conserved charges whose spins are
$s=\pm 1, \pm 2, \dots ,\pm r\ {\rm mod}\ (r+1)$. Spin $\pm 1$ corresponds to
the energy-momentum vector, and the conserved quantities of other spins
correspond
to other conserved tensors of higher rank. It is expected
half of these (energy-like), at most, could be conserved on the half-line.

For the whole-line theory, it is convenient to think in terms of light-cone
coordinates and densities for spin $s$ satisfying,
$$\partial_\mp T_{\pm (s+1)}=\partial_\pm \Theta_{\pm (s-1)}.$$
However, on the half-line  the energy-like combinations are the
relevant candidates and the quantities $\widehat P_s$ (the spin label will
continue
to be used even though Lorentz invariance has been lost) defined by
\eqn\Ps{\widehat P_s=\int_{-\infty}^0dx\left(T_{s+1}+T_{-s-1}-
\Theta_{s-1}-\Theta_{-s+1}\right)-\Sigma_s(\phi_0 ),}
where the additional term must be chosen to satisfy
\eqn\Pcondition{T_{s+1}-T_{-s-1}+
\Theta_{s-1}-\Theta_{-s+1}={\partial\phi_a\over \partial t}
{\partial\Sigma_s\over \partial\phi_a}.}
Eq(\Pcondition ) is remarkably strong. Ghoshal and Zamolodchikov appear to
be the first to use such an argument for the case of the spin-three charge in
the
sine-Gordon model.\NRF\rGZa\GZa\refs{\rGZa}

For low spin charges, such as occur in the $a_n^{(1)}$ theories
($s=2$ for $n>1$), and
the $a_1^{(1)}$ and $d_n^{(1)}$ cases ($s=3$), it is
straightforward to examine
\Pcondition\ directly, and many of the details are available
elsewhere.\NRF\rCDRSa\CDRSa
\NRF\rBCDRa\BCDRa\refs{\rCDRSa ,\rBCDRa} The conclusion is the
following. For all of
these models, the boundary potential
must take the form
\eqn\boundary{{\cal B}=\sum_0^rA_ie^{\alpha_i\cdot\phi /2}}
where, {\bf either} every coefficient vanishes
(the Neumann condition) {\bf or},
every coefficient is non-zero with magnitude $2\sqrt{n_i}$, {\bf except} for
the case $a_1^{(1)}$, where the two coefficients are arbitrary.

It is tempting to conjecture that the form of the boundary potential provided
by \boundary\ is universal. This is indeed so, but the restrictions on
the coefficients are not quite applicable in every case. The Lax pair
approach reveals that the strong restrictions on the coefficients $A_i$
apply to every $ade$-type model but are not quite universal. The
second class of models, based on the non simply-laced algebras,
most occurring as dual pairs, allows a small amount
of freedom in the choice of boundary data (see the table reproduced at the end
of the section). However, it is
only in the sine-Gordon case that the maximum freedom is permitted. Note
also, in most cases, setting the field to a specific value at the boundary
will not be compatible with integrability in the sense described. In other
words,
Dirichlet boundary conditions are not generally permitted.

Actually, even in the sinh-Gordon case there is a question of stability.
Recall the Bogomolny bound argument and consider the total energy for
a time independent solution to the theory restricted to a half-line
\NRF\rCDRa\CDRa\refs{\rCDRa}:
$$\eqalign{\widehat E&=\int_{-\infty}^0 dx \left({1\over 2}(\phi^\prime )^2 +
e^{\sqrt{2}\phi}+e^{-\sqrt{2}\phi}-2\right) +A_1e^{\phi_0/\sqrt{2}}
+A_0 e^{-\phi_0/\sqrt{2}}\cr
&=\int_{-\infty}^0 dx {1\over 2}\left(\phi^\prime -\sqrt{2}e^{\phi /\sqrt{2}}
+\sqrt{2}e^{-\phi /\sqrt{2}}\right)^2 +\int_{-\infty}^0 dx \sqrt{2}
\phi^\prime\left(e^{\phi /\sqrt{2}}-e^{-\phi /\sqrt{2}}\right) + \dots\cr
&\ge -4 +(A_0+2)e^{-\phi_0 /\sqrt{2}} +(A_1+2)e^{\phi_0 /\sqrt{2}}.\cr}$$
It is clear that the energy is bounded below provided $A_0\ge -2$ and
$A_1\ge -2$. Further details on the question of stability in this and other
cases are to be found  in
Fujii and Sasaki.\NRF\rFSa\FSa\refs{\rFSa} There is a similar
Bogomolny style argument for the
$a_2^{(2)}$ (or Bullough-Dodd) model
\NRF\rCHa{\CHa\semi\Hh}\refs{\rCHa}. However, if there exists such an
argument for the other cases, it does not appear to be have been written down.

The form of \boundary\ has been discovered by examining low spin charges
but there is always the possibility that some higher spin charge will violate
integrability unless further, more stringent, conditions are imposed. To
ensure compatibility with infinitely many charges it will be necessary
to adopt a different approach and to develop the Lax pair idea beyond
its formulation for the whole line.

First, the basic idea of a Lax pair requires the discovery
of a \lq gauge field' whose curvature vanishes if and only if
the field equations for the fields $\phi_a$ are satisfied. Explicitly,
for affine Toda theory, the
Lax pair may be chosen to be,\NRF\rMOPa{\MOPa\semi\Wa\semi\OTa}\refs{\rMOPa}
\eqn\laxfull{\eqalign{&a_0=H\cdot\partial_1\phi /2+\sum_0^r
\sqrt{m_i}(\lambda E_{\alpha_i}-1/\lambda \ E_{-\alpha_i}) e^{\alpha_i\cdot\phi
/2}\cr
&a_1=H\cdot\partial_0\phi /2+\sum_0^r
\sqrt{m_i}(\lambda E_{\alpha_i}+1/\lambda \ E_{-\alpha_i}) e^{\alpha_i\cdot\phi
/2},\cr}}
where $H_a, E_{\alpha_i}$ and $E_{-\alpha_i}$ are the Cartan subalgebra
and the generators
corresponding to the simple roots, respectively, of the simple Lie algebra
providing the data for the Toda theory.
The coefficients $m_i$ are related to the $n_i$ by $m_i=n_i \alpha_i^2/8$.
The conjugation properties of the generators are chosen so that
\eqn\conj{a_1^\dagger  (x,\lambda )=a_1  (x,1/\lambda )
\qquad a_0^\dagger  (x,\lambda )
=a_0 (x,-1/\lambda ).}
Using the Lie algebra relations
$$[H, E_{\pm\alpha_i}]=\pm\, \alpha_i\, E_{\pm \alpha_i}\qquad
[E_{\alpha_i},E_{-\alpha_i}]=
2\alpha_i\cdot H/(\alpha_i^2),$$
the zero curvature condition for \laxfull\
$$f_{01}=\partial_0a_1-\partial_1a_0 +[a_0,a_1]=0$$
leads to the affine Toda field equations:
\eqn\todafull{\partial^2\phi =-\sum_0^r n_i \alpha_i e^{\alpha_i\cdot\phi}.}
\medskip
\noindent{\bf Modified Lax pair}
\medskip
For the purposes of the following discussion (which follows very closely the
article by Bowcock et al.\NRF
\rBCDRa\BCDRa\refs{\rBCDRa}) the boundary of the half-line
will be placed at $x=a$.

To construct a modified Lax pair including the boundary condition
derived from \Lboundary ,
it was found in  to be convenient to
consider an additional  special point $x =b\ (>a)$ and two overlapping
regions $R_-:\ x \le (a+b+\epsilon )/2;\ $ and $R_+:\ x \ge (a+b-\epsilon
)/2$.
The second region will be regarded as a reflection of the first,
in the sense that if $x \in R_+$, then
\eqn\reflectphi{\phi (x )\equiv\phi (a+b-x ).}
The regions overlap in a small interval surrounding the midpoint of $[a,b]$.
In the two regions define:
\eqn\newlax{\eqalign{&R_-:\qquad \widehat a_0=a_0 -{1\over 2}\theta (x -a)
\left(\partial_1\phi +
{\partial\B\over\partial\phi}\right)\cdot H \qquad
\widehat a_1=\theta (a-x )a_1\cr
&R_+:\qquad \widehat a_0=a_0 -{1\over 2}\theta (b-x )
\left(\partial_1\phi -
{\partial\B\over\partial\phi}\right)\cdot H \qquad
\widehat a_1=\theta (x -b)a_1.\cr}}
Then, it is clear that in the region $x <a$ the Lax pair \newlax\ is
the same as the old but, at $x =a$ the derivative of
the $\theta$ function in the zero curvature condition enforces the boundary
condition
\eqn\boundary{{\partial\phi\over\partial x }=-{\partial{\cal B}\over \partial
\phi}, \qquad x =a .}
Similar statements hold for $x \ge b$
except that the
boundary condition at $x =b$ differs by a sign  in order to
accommodate the reflection condition \reflectphi .

On the other hand, for $x \in R_-$ and $x >a$, $\widehat a_1$ vanishes
and therefore the zero curvature condition merely implies $\widehat a_0$
is independent of $x $. In turn, this fact implies  $\phi$ is
independent of $x $ in this region. Similar remarks apply to the region
$x \in R_+$ and $x <b$. Hence, taking into account the reflection principle
\reflectphi , $\phi$ is independent of $x $ throughout the interval $[a,b]$,
and equal to its value at $a$ or $b$. For general boundary conditions, a glance
at
\newlax\ reveals that the gauge potential $\widehat a_0$ is different in the
two
regions $R_\pm$. However, to maintain the zero curvature condition over the
whole
line the values of $\widehat a_0$ must be related by a gauge transformation
on the overlap. Since $\widehat a_0$ is in fact independent of $x \in [a,b]$
on both patches, albeit  with a different value on each patch,
the zero curvature condition effectively requires the existence of
a gauge transformation $\K$ with the property:
\eqn\Kdef{\partial_0 \K =\K\, \widehat a_0(t ,b) -\widehat a_0(t ,a)\, \K .}
The group element $\K$ lies in the group $G$ with Lie algebra $g$, the
Lie algebra whose roots define the  affine Toda theory.

The conserved quantities on the half-line ($x\le a$) are
determined via a generating function $\widehat Q(\lambda )$ given by the
expression
\eqn\Qalt{\widehat Q(\lambda )={\rm tr}\left( U(-\infty ,a;\lambda )\
\K \ U^\dagger(-\infty, a ;1/\lambda )\right),}
where $U(x_1,x_2;\lambda )$ is defined by the path-ordered exponential:
\eqn\pathexp{U(x_1,x_2;\lambda )={\rm P}\exp \int_{x_1}^{x_2} dx\,  a_1 .}

To further understand the nature of $\K$, it is convenient to make a couple of
assumptions which turn out to be no more restrictive as far as
the boundary potential is concerned than the investigation of the low spin
charges. Suppose $\K$ is time independent, and also independent of $\phi$
in a functional sense. Then, \Kdef\ simplifies to
$$\K\, \widehat a_0(t ,b) -\widehat a_0(t ,a)\, \K =0$$
or, in terms of the explicit expression for $\widehat a_0$,
\eqn\Kdefa{{1\over 2}\left[\K (\lambda ),\,
{\partial\B\over\partial\phi}\cdot H\right]_+=-\,\left[\K (\lambda )
,\, \sum_0^r
\sqrt{m_i}(\lambda E_{\alpha_i}-1/\lambda \ E_{-\alpha_i})
e^{\alpha_i\cdot\phi /2}\right]_-,}
where the field-dependent quantities are evaluated at the boundary $x=a$.
Eq\Kdefa\ is very strong, not only determining ${\cal B}$ but also $\K$
almost uniquely. The details of many solutions, including a catalogue of the
restrictions on ${\cal B}$ have been found
\NRF\rBCDRa\BCDRa\refs{\rBCDRa}. Here, just
two examples will be given for $\K$, and a list of the parameter
restrictions for ${\cal B}$. For $\K$ the overall scale is a matter of
convenience only.

$$\eqalign{a_1^{(1)}&:\quad
\K (\lambda )=\left(\lambda^2-{1\over\lambda^2}\right){\rm I} + \pmatrix{&0&
\lambda A_1-{A_0\over\lambda }\cr
&\lambda A_0-{A_1\over\lambda} &0\cr}\cr
&\ \ \cr
a_n^{(1)}&:\quad \K (\lambda )=I+2\sum_{\alpha >0}\prod_i C_i^{l_i(\alpha )}
\left[{(-\lambda)^{l(\alpha )}E_\alpha\over1+C\lambda^h}+{(-1/\lambda
)^{l(\alpha )}
E_{-\alpha}\over 1+C/\lambda^{-h}}\right],\cr}$$
where, in the latter expression, $C_i=A_i/2$, $C=\prod_iC_i$, each positive
root in
the sum may be decomposed as a sum of simple roots and $l_i(\alpha )$ denotes
the
number of times $\alpha_i$ appears in the decomposition, $l(\alpha )=
\sum_i l_i(\alpha )$.

As far as the boundary potential is concerned, the conjecture mentioned above
appears to be correct for the $ade$ series of models, implying the strongly
restricted boundary parameters. For all the others, the form of the boundary
potential is the same but the restrictions on the parameters are less severe.
The  diagrams below represent the possibilities. The symbols next to the
circles representing simple roots indicate the type of coefficient the
corresponding term may have in the boundary potential. Where there is only a
discrete
choice, it is labelled by $\epsilon$, denoting $\pm$, and any choice of signs
is permitted. These would be the only labels for the $ade$ diagrams.
Where there is a continuous paramter associated with a term,
it is labelled by $x$ or $y$. The labels above the circles represent
one set
of possibilities, those below represent an alternative set;
the Neumann condition is possible in all cases (but not Dirichlet). In very
few cases is there the possibility of continuously deforming away from the
Neumann condition while maintaining integrability.

Finally, once $\K (\lambda )$ is determined, it is necessary to demonstrate its
compatibility with the classical $r$ matrix which itself determines the
Poisson brackets between the generating functions for the conserved charges
defined for the whole line theory (see for example the articles by Olive and
Turok
\NRF\rOTb{\OTb\semi\OTd}\refs{\rOTb}).

Explicitly,
$$\{ U(\lambda ){\otimes ,} U(\mu )\}=\left[ r(\lambda /\mu ),
U(\lambda )\otimes U(\mu )\right],$$
where $r$ has the form
$$r(s)=\sum_ir_i(s)g_i\otimes g_i^\dagger ,$$
and
$U(\lambda )$ is defined in \pathexp . The quantities $g_i$ represent the Lie
algebra generators for an algebra whose root system defines
a particular Toda model.
Calculating the Poisson brackets between two charges of the form
given by \Qalt , will clearly require a consistency condition to be satisfied
involving $r$ and $\K$.
In earlier work,\NRF\rSl{\Sl\semi\Sm}\refs{\rSl} the compatibility relation
appears as the main equation to be satisfied by $\K (\lambda )$ whereas
here, $\K$ has been determined independently via Eq.\Kdefa .

The necessary
checking
has been carried out
\NRF\rBCDRa\BCDRa\refs{\rBCDRa} and $\K$ is indeed compatible
with $r$. In other words, it satisfies the following:

$$\left[r(\lambda /\mu ),\K^{(1)}(\lambda )\K^{(2)}(\mu )\right]
=\K^{(1)}(\lambda )\tilde{r}(\lambda\mu )\K^{(2)}(\mu )-\K^{(2)}(\mu )
\tilde{r}(\lambda\mu )
\K^{(1)}(\lambda ),$$
where
$$\K^{(1)}(\lambda )=\K (\lambda )\otimes 1,\qquad \K^{(2)}(\mu )=1\otimes
\K (\mu ),$$
and
$$\tilde{r}(s )=\sum_ir_i(s )g_i\otimes g_i.$$
The relationship between $r$ and $\K$ is one which would probably
repay further study; in a sense $\K$ is a fundamental object and
presumably one could argue that the classical $r$ matrix must be
chosen to be compatible with it. Indeed it appears remarkable that
$\K$ and $r$ are compatible given they have been determined
independently, and given the seemingly strong
assumptions made to derive the expressions for $\K$ in
the various examples. Even in the quantum
case,
as discussed earlier, there is a set of reflection bootstrap
equations which would actually allow a computation of the complete set of
S-matrix factors
if the full set
of reflection factors had  been already independently determined.
\vfill\eject

\epsffile{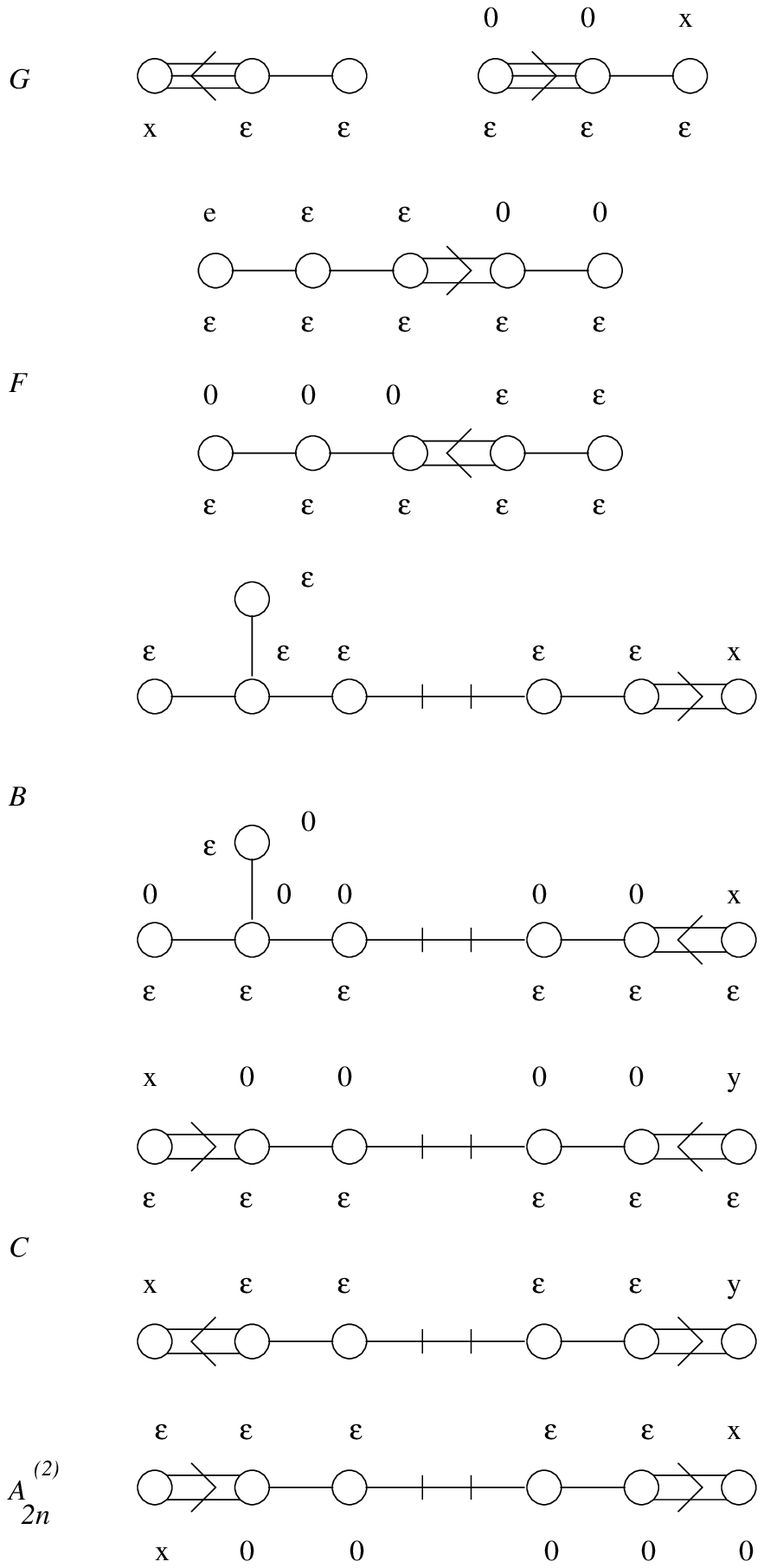}

\newsec{Classical reflection factors}

Following the programme described earlier, the next step is to discover
the classical reflection factors for each of the permissible boundary
conditions. This is an unfinished part of the programme but at least
a  method for carrying it out can be relatively easily, and briefly,
described. First, one needs to know the lowest energy static solution
for a given theory with a specific boundary condition. In general,
this solution will not be $\phi_a=0$.
These are, in
effect stationary \lq solitons' in the following sense. The real
affine Toda theories cannot have finite energy localised solutions because
all static solutions diverge somewhere. Nevertheless, where there is a
boundary these solutions are of relevance because the divergence can be
placed harmlessly on the other side of it. I.e. for $x<0$
there are non-singular solutions each of which is a portion of a
\lq soliton'. Similarly, on a finite interval,
periodic solutions may be used whose singularities lie outside the
interval.

The simplest example of this is the $\sinh$-Gordon
model.\NRF\rCDRa\CDRa\refs{\rCDRa} There,
the equation for the static background is

\eqn\sg{\eqalign{\phi ^{\prime\prime}&=-\sqrt{2}\left(e^{\sqrt{2}\phi }-
e^{-\sqrt{2}\phi }\right)\qquad\qquad\qquad x<0\cr
\phi ^\prime&=-\sqrt{2}\left(\epsilon_1e^{\phi  /\sqrt{2}}-\epsilon_0e^{-\phi
/\sqrt{2}}\right)\qquad x=0, \qquad A_i=2\epsilon_i\cr}}
from which, on integrating the first equation once, and comparing with
the boundary condition, one obtains
\eqn\sga{\eqalign{\phi ^\prime& =\sqrt{2}\left(e^{\phi  /\sqrt{2}}-
e^{-\phi  /\sqrt{2}}\right)\qquad x<0\cr
e^{\sqrt{2}\phi }&={1+\epsilon_0\over 1+\epsilon_1}\qquad
\qquad\qquad\qquad x=0.\cr}}
Hence, the static solution takes the form
\eqn\sgbackground{e^{\phi  /\sqrt{2}}={1+e^{2(x-x_0)}\over
1-e^{2(x-x_0)}},}
with
\eqn\sgbv{\hbox{coth}\, x_0=\sqrt{1+\epsilon_0\over 1+\epsilon_1}.}
The expression given in \sgbv\ assumes $\epsilon_0>\epsilon_1$;
if that is not the case, it is necessary to adjust the solution by
shifting $x_0$ by $i\pi /2$. Provided $x_0$ is positive the singularity
in Eq.\sgbackground\ is irrelevant.

Other examples for the series $a_n^{(1)}$ have been calculated by
Bowcock.\NRF\rBa\Ba\refs{\rBa}

Once the static background is known the classical reflection factors
are sought by linearising the field equation and the boundary condition,
and calculating the reflection of a plane wave in the effective potential
due to the static background.

Thus, within the $\sinh$-Gordon model the classical background is given by
Eq.\sgbackground\ and
the linearised wave equation and boundary condition
in this background have the form
\eqn\sglin{\eqalign{\partial^2\phi^{(0)}&=-4\left(1+{2\over
\sinh^2 2(x-x_0)}\right)\phi^{(0)}\qquad x<0\cr
\partial_1\phi^{(0)} &=-\left(\epsilon_0\tanh x_0 +\epsilon_1
\coth x_0\right)\phi^{(0)}\qquad x=0.\cr}}
The solitonic nature of the static solution is now fairly
evident since it leads to
a \lq sech$^2$' effective potential which is well-known to be
exactly solvable;
this potential is  also known to be related to solitons in various
ways.\NRF\rLa\La\refs{\rLa}

The classical scattering data for this potential
is computable  in terms
of the parameters in the boundary term. It is convenient to set
$\phi^{(0)}=e^{-i\omega t}\Phi(z)$, in which case the solution to \sglin\
takes the form
$$\Phi (z)=a(i\lambda -\coth (z-z_0))e^{i\lambda z} +cc,\qquad
\lambda =\sinh\theta ,$$
where the ratio of coefficients $a^*/a$ can be computed from the boundary
condition.  The reflection coefficient may be read off and turns out to be
\eqn\sgcoeff{\eqalign{K&= {1-i\lambda \over 1+i\lambda}\, {(i\lambda)^2+
i\lambda\sqrt{1+\epsilon_0}\sqrt{1+\epsilon_1}+(\epsilon_0+\epsilon_1)/2
\over (i\lambda)^2-
i\lambda\sqrt{1+\epsilon_0}\sqrt{1+\epsilon_1}+(\epsilon_0+\epsilon_1)/2}\cr
&=-\,
(1)^2~\left[(1+a_0+a_1)(1-a_0+a_1)(1+a_0-a_1)(1-a_0-a_1)\right]^{-1},\cr}}
where in the last step it has been convenient to set
$$\epsilon_i=\cos a_i\pi , \qquad |a_i|\le 1,\qquad i=0,1\ ,$$
and to use the notation
\eqn\bracket{(z)={\sinh \left({\theta\over 2} +{i\pi z\over 2h}\right)\over
\sinh \left({\theta\over 2} -{i\pi z\over 2h}\right) }\qquad h=2.}
To extend beyond the restriction on the $a_i$, it is necessary to continue
the formula \sgcoeff\ by making the substitution $a_i\rightarrow a_i+2$.
The result Eq.\sgcoeff\ is remarkably similar to the conjectured quantum
reflection coefficient for the lightest sine-Gordon breather which
has been provided by Ghoshal.\NRF\rGd\Gd\refs{\rGd} However, it remains
unclear
how the parameters of the classical boundary potential are related to the
parameters in his conjecture.

A similar calculation may be made in other cases. However, a more general
procedure is the following.\NRF\rCDRa\CDRa\NRF\rBa\Ba\refs{\rCDRa ,\rBa}
Once it is realised that the background is a \lq soliton', the
corresponding \lq multisolitons' must provide solutions in which solitons
scatter from the boundary. In particular, a triple, or possibly
other multi-soliton, solution
containing the static background as one of its components, should, in
a suitable limit, directly provide the classical reflection factors.
At least in priciple, all the multi-soliton solutions are known because
they ought to be simply related to the complex solitons
mentioned before.\refs{\rHa} From these, the classical reflection data
can be determined.

Once the classical reflection data has been determined it should be
possible to check the classical version of the bootstrap condition.

For example, in the $d_5^{(1)}$ case, with the following configuration of
boundary coefficients ($a=-2$ and $b=-2\sqrt{2}$):
\vfill\eject
\epsffile{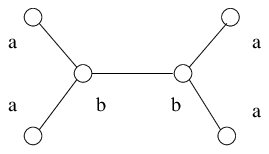}
\vskip -3.5cm
\noindent the classical static soliton is given by
\eqn\dfivesol{\phi =(\alpha_2 +\alpha_3 )\psi, \qquad e^{\psi /2}=
{1-e^{2x}\over 1+e^{2x}}\ ,}
where $\alpha_2$ and $\alpha_3$ are the simple roots corresponding to
the two central circles of the above Dynkin-Ka\v {c} diagram. Notice
the solution actually diverges just at the boundary; nevertheless, its
energy is finite because of a cancellation between the integral of
the energy density over the half-line and the boundary potential.

The calculation of the classical reflection factors is quite complicated
and will not be reproduced here. However, the result is the following:
\eqn\dfiverf{\K_s=\K_{\bar s}=-(2)(6)\qquad \K_1={(3)(5)\over (1)(7)}\qquad
\K_2=-(4)^2\qquad \K_3=(1)(3)(5)(7),}
where the notation of Eq.\bracket\ has been used with $h=8$
(the Coxeter number of $d_5$), and where the particle labels are
the standard ones for affine Toda field theory.\NRF\rBCDSc\BCDSc\refs{\rBCDSc}
(The labels $s$ and $\bar s$ are reserved for the particles corresponding
to one of the forks on the diagram above; $2$ and $3$ label the particles
associated with the centre circles; particle 1 is associated with one of the
circles of the other fork, and the final circle corresponds to the extended
affine root and has no particle associated with it.) This particular model has
three-point couplings:
$$ ss1\quad ss3\quad s\bar s 2\quad 112\quad 123\quad 332$$
and it is not hard to check that the bootstrap associated with these couplings
is satisfied by the classical reflection factors given in Eq.\dfiverf .

\newsec{\bf Acknowledgements}

I would like to thank the organisers of the Conference and the Physics
Department at Xinjiang University for their
kind hospitality and the opportunity to speak about these topics.  I am
indebted
to Medina Ablikim, Peter Bowcock, Patrick Dorey, Rachel Rietdijk and
Ryu Sasaki for many illuminating discussions.

\vfill\eject
\footatend\vfill\supereject\immediate\closeout\rfile\writestoppt
\baselineskip=12pt\ni{\bf References}\medskip{\frenchspacing%
\parindent=20pt\escapechar=` \input refs.tmp\vfill\eject}\nonfrenchspacing

\end

%% file: wsharvmac.tex
%
%
%
%
%
\def\unredoffs{\hoffset-.14truein\voffset-.2truein} 
\def\redoffs{\voffset=-.45truein\hoffset=-.21truein} 
\def\speclscape{}
%
\def\ni{\noindent}
\newbox\leftpage \newdimen\fullhsize \newdimen\hstitle \newdimen\hsbody
\tolerance=1000\hfuzz=2pt
\catcode`\@=11 
\def\bigans{b }
\let\answ=\bigans
\ifx\answ\bigans\message{(This will come out unreduced.}
\magnification=1200\unredoffs\baselineskip=12pt plus 2pt minus 1pt
\parskip=0pt 
\hoffset=.3truein \voffset=.2truein \vsize=9truein
\hsbody=6.3truein \hsize=\hsbody \hstitle=\hsize 
\else\message{(This will be reduced.} \let\l@r=L
\magnification=1000\baselineskip=16pt plus 2pt minus 1pt \vsize=9truein
\redoffs \hstitle=8truein\hsbody=6.3truein\fullhsize=10truein\hsize=\hsbody
\output={\ifnum\pageno=0 
  \shipout\vbox{\speclscape{\hsize\fullhsize\makeheadline}
    \hbox to \fullhsize{\hfill\pagebody\hfill}}\advancepageno
  \else
  \almostshipout{\leftline{\vbox{\pagebody\makefootline}}}\advancepageno 
  \fi }
\def\almostshipout#1{\if L\l@r \count1=1 \message{[\the\count0.\the\count1]}
      \global\setbox\leftpage=#1 \global\let\l@r=R
 \else \count1=2
  \shipout\vbox{\speclscape{\hsize\fullhsize\makeheadline}
      \hbox to\fullhsize{\box\leftpage\hfil#1}}  \global\let\l@r=L\fi}
\fi
%
\newcount\yearltd\yearltd=\year\advance\yearltd by -1900

%
%

\def\draftmode{\message{ DRAFTMODE }\def\draftdate{{\rm preliminary draft:
\number\month/\number\day/\number\yearltd\ \ \hourmin}}%
\headline={\hfil\draftdate}\writelabels\baselineskip=20pt plus 2pt minus 2pt
 {\count255=\time\divide\count255 by 60 \xdef\hourmin{\number\count255}
  \multiply\count255 by-60\advance\count255 by\time
  \xdef\hourmin{\hourmin:\ifnum\count255<10 0\fi\the\count255}}}
\def\nolabels{\def\wrlabeL##1{}\def\eqlabeL##1{}\def\reflabeL##1{}}
\def\writelabels{\def\wrlabeL##1{\leavevmode\vadjust{\rlap{\smash%
{\line{{\escapechar=` \hfill\rlap{\sevenrm\hskip.03in\string##1}}}}}}}%
\def\eqlabeL##1{{\escapechar-1\rlap{\sevenrm\hskip.05in\string##1}}}%
\def\reflabeL##1{\noexpand\llap{\noexpand\sevenrm\string\string\string##1}}}
\nolabels
%
\global\newcount\secno \global\secno=0
\global\newcount\meqno \global\meqno=1
\def\newsec#1{\global\advance\secno by1\message{(\the\secno. #1)}
\global\subsecno=0\eqnres@t\noindent{\bf\the\secno. #1}
\writetoca{{\secsym} {#1}}\par\nobreak\medskip\nobreak}
\def\eqnres@t{\xdef\secsym{\the\secno.}\global\meqno=1\bigbreak\bigskip}
\def\sequentialequations{\def\eqnres@t{\bigbreak}}\xdef\secsym{}
\global\newcount\subsecno \global\subsecno=0
\def\subsec#1{\global\advance\subsecno by1\message{(\secsym\the\subsecno. #1)}
\ifnum\lastpenalty>9000\else\bigbreak\fi
\noindent{\it\secsym\the\subsecno. #1}\writetoca{\string\quad 
{\secsym\the\subsecno.} {#1}}\par\nobreak\medskip\nobreak}
\def\appendix#1#2{\global\meqno=1\global\subsecno=0\xdef\secsym{\hbox{#1.}}
\bigbreak\bigskip\noindent{\bf Appendix #1. #2}\message{(#1. #2)}
\writetoca{Appendix {#1.} {#2}}\par\nobreak\medskip\nobreak}
%
\newcount\qno   \qno=1
\def\Q{\item{[\the\qno ]\ }\advance\qno by +1}
%
%
\def\eqnn#1{\xdef #1{(\secsym\the\meqno)}\writedef{#1\leftbracket#1}%
\global\advance\meqno by1\wrlabeL#1}
\def\eqna#1{\xdef #1##1{\hbox{$(\secsym\the\meqno##1)$}}
\writedef{#1\numbersign1\leftbracket#1{\numbersign1}}%
\global\advance\meqno by1\wrlabeL{#1$\{\}$}}
\def\eqn#1#2{\xdef #1{(\secsym\the\meqno)}\writedef{#1\leftbracket#1}%
\global\advance\meqno by1$$#2\eqno#1\eqlabeL#1$$}
%
\newskip\footskip\footskip14pt plus 1pt minus 1pt 
\def\footnotefont{\ninepoint}\def\f@t#1{\footnotefont #1\@foot}
\def\f@@t{\baselineskip\footskip\bgroup\footnotefont\aftergroup\@foot\let\next}
\setbox\strutbox=\hbox{\vrule height9.5pt depth4.5pt width0pt}
\global\newcount\ftno \global\ftno=0
\def\foot{\global\advance\ftno by1\footnote{$^{\the\ftno}$}}
%
\newwrite\ftfile   
\def\footend{\def\foot{\global\advance\ftno by1\chardef\wfile=\ftfile
$^{\the\ftno}$\ifnum\ftno=1\immediate\openout\ftfile=foots.tmp\fi%
\immediate\write\ftfile{\noexpand\smallskip%
\noexpand\item{f\the\ftno:\ }\pctsign}\findarg}%
\def\footatend{\vfill\eject\immediate\closeout\ftfile{\parindent=20pt
\centerline{\bf Footnotes}\nobreak\bigskip\input foots.tmp }}}
\def\footatend{}
%
%
\global\newcount\refno \global\refno=1
\newwrite\rfile
\def\ref{[\the\refno]\nref}
\def\nref#1{\xdef#1{[\the\refno]}\writedef{#1\leftbracket#1}%
\ifnum\refno=1\immediate\openout\rfile=refs.tmp\fi
\global\advance\refno by1\chardef\wfile=\rfile\immediate
\write\rfile{\noexpand\item{#1\ }\reflabeL{#1\hskip.31in}\pctsign}\findarg}
\def\findarg#1#{\begingroup\obeylines\newlinechar=`\^^M\pass@rg}
{\obeylines\gdef\pass@rg#1{\writ@line\relax #1^^M\hbox{}^^M}%
\gdef\writ@line#1^^M{\expandafter\toks0\expandafter{\striprel@x #1}%
\edef\next{\the\toks0}\ifx\next\em@rk\let\next=\endgroup\else\ifx\next\empty%
\else\immediate\write\wfile{\the\toks0}\fi\let\next=\writ@line\fi\next\relax}}
\def\striprel@x#1{} \def\em@rk{\hbox{}} 
\def\lref{\begingroup\obeylines\lr@f}
\def\lr@f#1#2{\gdef#1{\ref#1{#2}}\endgroup\unskip}
\def\semi{;\hfil\break}
\def\addref#1{\immediate\write\rfile{\noexpand\item{}#1}} 
\def\startrefs#1{\immediate\openout\rfile=refs.tmp\refno=#1}
\def\xref{\expandafter\xr@f}\def\xr@f[#1]{#1}
\def\refs#1{\count255=1\hskip -3pt $^{\r@fs#1{\hbox{}}}$}
\def\r@fs#1{\ifx\und@fined#1\message{reflabel \string#1 is undefined.}%
\nref#1{need to supply reference \string#1.}\fi%
\vphantom{\hphantom{#1}}\edef\next{#1}\ifx\next\em@rk\def\next{}%
\else\ifx\next#1\ifodd\count255\relax\xref#1\count255=0\fi%
\else#1\count255=1\fi\let\next=\r@fs\fi\next}
%

%
\newwrite\ffile\global\newcount\figno \global\figno=1
\def\fig{fig.~\the\figno\nfig}
\def\nfig#1{\xdef#1{fig.~\the\figno}%
\writedef{#1\leftbracket fig.\noexpand~\the\figno}%
\ifnum\figno=1\immediate\openout\ffile=figs.tmp\fi\chardef\wfile=\ffile%
\immediate\write\ffile{\noexpand\medskip\noexpand\item{Fig.\ \the\figno. }
\reflabeL{#1\hskip.55in}\pctsign}\global\advance\figno by1\findarg}
\def\xfig{\expandafter\xf@g}\def\xf@g fig.\penalty\@M\ {}
\def\figs#1{figs.~\f@gs #1{\hbox{}}}
\def\f@gs#1{\edef\next{#1}\ifx\next\em@rk\def\next{}\else
\ifx\next#1\xfig #1\else#1\fi\let\next=\f@gs\fi\next}
\newwrite\lfile
{\escapechar-1\xdef\pctsign{\string\%}\xdef\leftbracket{\string\{}
\xdef\rightbracket{\string\}}\xdef\numbersign{\string\#}}

\def\writestop{\def\writestoppt{\immediate\write\lfile{\string\pageno%
\the\pageno\string\startrefs\leftbracket\the\refno\rightbracket%
\string\def\string\secsym\leftbracket\secsym\rightbracket%
\string\secno\the\secno\string\meqno\the\meqno}\immediate\closeout\lfile}}
\def\writestoppt{}\def\writedef#1{}
\def\seclab#1{\xdef #1{\the\secno}\writedef{#1\leftbracket#1}\wrlabeL{#1=#1}}
\def\subseclab#1{\xdef #1{\secsym\the\subsecno}%
\writedef{#1\leftbracket#1}\wrlabeL{#1=#1}}
\newwrite\tfile \def\writetoca#1{}
\def\leaderfill{\leaders\hbox to 1em{\hss.\hss}\hfill}
\def\writetoc{\immediate\openout\tfile=toc.tmp 
   \def\writetoca##1{{\edef\next{\write\tfile{\noindent ##1 
   \string\leaderfill {\noexpand\number\pageno} \par}}\next}}}

%
\catcode`\@=12 
%
\edef\tfontsize{\ifx\answ\bigans scaled\magstep3\else scaled\magstep4\fi}
 \tfontsize  \tfontsize
 \tfontsize \font\titlei=cmmi10 \tfontsize
\font\titleis=cmmi7 \tfontsize \font\titleiss=cmmi5 \tfontsize
\font\titlesy=cmsy10 \tfontsize \font\titlesys=cmsy7 \tfontsize
\font\titlesyss=cmsy5 \tfontsize  \tfontsize
\skewchar\titlei='177 \skewchar\titleis='177 \skewchar\titleiss='177
\skewchar\titlesy='60 \skewchar\titlesys='60 \skewchar\titlesyss='60
 \ifx\answ\bigans\else scaled\magstep1\fi
\ifx\answ\bigans\else

 \font\absi=cmmi10 scaled\magstep1
\font\absis=cmmi7 scaled\magstep1 \font\absiss=cmmi5 scaled\magstep1
\font\abssy=cmsy10 scaled\magstep1 \font\abssys=cmsy7 scaled\magstep1
\font\abssyss=cmsy5 scaled\magstep1 
\skewchar\absi='177 \skewchar\absis='177 \skewchar\absiss='177
\skewchar\abssy='60 \skewchar\abssys='60 \skewchar\abssyss='60
\fi
\font\ninerm=cmr9 \font\sixrm=cmr6 \font\ninei=cmmi9 \font\sixi=cmmi6 
\font\ninesy=cmsy9 \font\sixsy=cmsy6 \font\ninebf=cmbx9 
\font\nineit=cmti9 \font\ninesl=cmsl9 \skewchar\ninei='177
\skewchar\sixi='177 \skewchar\ninesy='60 \skewchar\sixsy='60 
\def\ninepoint{\def\rm{\fam0\ninerm}
\textfont0=\ninerm \scriptfont0=\sixrm \scriptscriptfont0=\fiverm
\textfont1=\ninei \scriptfont1=\sixi \scriptscriptfont1=\fivei
\textfont2=\ninesy \scriptfont2=\sixsy \scriptscriptfont2=\fivesy
\textfont\itfam=\ninei \def\it{\fam\itfam\nineit}\def\sl{\fam\slfam\ninesl}%
\textfont\bffam=\ninebf \def\bf{\fam\bffam\ninebf}\rm} 
%
%
\def\noblackbox{\overfullrule=0pt}
\hyphenation{anom-aly anom-alies coun-ter-term coun-ter-terms}
\def\inv{^{\raise.15ex\hbox{${\scriptscriptstyle -}$}\kern-.05em 1}}

\def\Dsl{\,\raise.15ex\hbox{/}\mkern-13.5mu D} 
\def\dsl{\raise.15ex\hbox{/}\kern-.57em\partial}

\def\lspace{\ifx\answ\bigans{}\else\qquad\fi}
\def\lbspace{\ifx\answ\bigans{}\else\hskip-.2in\fi} 
\def\boxeqn#1{\vcenter{\vbox{\hrule\hbox{\vrule\kern3pt\vbox{\kern3pt
	\hbox{${\displaystyle #1}$}\kern3pt}\kern3pt\vrule}\hrule}}}
\def\mbox#1#2{\vcenter{\hrule \hbox{\vrule height#2in
		\kern#1in \vrule} \hrule}}  
%

\def\darr#1{\raise1.5ex\hbox{$\leftrightarrow$}\mkern-16.5mu #1}

\def\roughly#1{\raise.3ex\hbox{$#1$\kern-.75em\lower1ex\hbox{$\sim$}}}